\documentclass[letterpaper,pre,showpacs,preprintnumbers,amsmath,amssymb]{revtex4}
      \newcommand{\beq}{\begin{equation}}
      \newcommand{\eeq}{\end{equation}}
      \newcommand{\beqa}{\begin{eqnarray}}
      \newcommand{\eeqa}{\end{eqnarray}}
      
      \newcommand{\nn}{\nonumber}

      \newcommand{\bra}{\left\langle}
      \newcommand{\ket}{\right\rangle}

      \newcommand{\al}{\alpha}
      \newcommand{\be}{\beta}
      
      \newcommand{\de}{\delta}

      \newcommand{\De}{\Delta}
      \newcommand{\La}{\Lambda}
      
      \newcommand{\bphi}{\mbox{\boldmath $\phi$}}

    \newcommand{\kt}{{\cal K}}
    \newcommand{\kl}{{\cal K}_<}
    \newcommand{\kh}{{\cal K}_>}
     \newcommand{\rZ}{{\cal Z}}
     \newcommand{\rH}{{\cal H}}
     \newcommand{\proj}{{\cal P}_0}
    \renewcommand{\(}{\left(}
    \renewcommand{\)}{\right)}

      \newcommand{\loc}{f(K)}

\newcommand{\rnd}[1]{{\mathbb E\, }{#1}}
\newcommand{\cum}[2]{{\kappa_{#1}\, }{\left(#2\right)}}

\usepackage{graphicx}
\usepackage{dcolumn}
\usepackage{bm}
\usepackage{mathrsfs}
\begin{document}
\preprint{}
\title
{
  Exactness of the replica method in perturbation 
  }
 
 \author{Hisamitsu Mukaida}
 \email{mukaida@saitama-med.ac.jp}
 \affiliation{Department of Physics, Saitama Medical University, 
 38 Moro hongo, Moroyama-cho, Iruma-gun, Saitama, 350-0495, Japan
 }
 
\author{Yoshinori Sakamoto}
\email{yossi@phys.ge.cst.nihon-u.ac.jp}
 \affiliation{Laboratory of Physics, College of Science and Technology, Nihon University,
   7-24-1, Narashino-dai, Funabashi-city, Chiba, 274-8501, Japan
 }

\date{090112 v1}

\begin{abstract}
The replica method for a quenched disordered system 
is considered in a perturbative field theory. 
Since correction in a finite-order perturbation is given in a polynomial of 
the replica number $n$, the zero-replica limit $n \rightarrow 0$ is regarded 
as extracting the constant term from the polynomial, which mathematically 
 makes sense. The meaning of the extraction is clarified comparing with 
a direct calculation. 
\end{abstract}


\maketitle
\section{Introduction}
The replica method is widely used for studying quenched disordered systems \cite{ea}. 
It was developed for getting around difficulty involved in taking  
average $\rnd{}(\cdot)$ over quenched disorder.   
 For example, 
in the case of a free energy,  we compute $\rnd{Z^n}$ instead of $\rnd{\log Z}$, 
where $Z$ is the partition function of the system computed 
under fixed quenched disorder, and use the formula
$$
  \rnd{\log Z} = \lim_{n \rightarrow 0} \frac{1}{n}(\rnd{Z^n} -1). 
$$
Here, $Z^n$ is realized by introducing $n$ ``replicants"
identical with the original system, so that $n$ is  
a positive integer, and thus the limit $n \rightarrow 0$ is ill-defined. 
Although there are specific models in which the replica method is proven to be
mathematically rigorous \cite{cfs,kmp,k},  justification for the zero-replica limit 
is left beyond the scope of investigation in most literatures.   
Nevertheless, since this method gives reliable results to various 
quenched disordered  systems, it has survived for more than three decades. 
In this paper, we wish to point out that the replica method can be justified within 
the framework of perturbation. 

The main idea for the justification is that a physical quantity in the replicated system 
calculated in a finite-order perturbation is a (finite-degree) polynomial in $n$. 
The zero-replica limit can be regarded as the extraction of the constant term from the 
polynomial. 
 A similar idea is used by Brunet and Derrida in Ref.\cite{b-d}. 

Generally, if a function $f(n)$ defined on positive integers 
 is a polynomial with the degree $m$,  
the constant term is obtained solving a linear equation  
for all the coefficients of $f(n)$ generated by $f(1), ..., f(m+1)$. 
We thus find that the constant term is extracted by 
\beq
  \sum_{n=1}^{m+1} (-1)^{n-1} {m+1 \choose n} f(n) \equiv \proj f(n). 
  \label{proj}
\eeq

It should be noted that $\proj$ in (\ref{proj}) becomes ill-defined  
when $m = \infty$, which implies that the constant term of a power series in $n$
becomes ambiguous. 
For instance,  consider the series 
$\sum_{k=0}^\infty (-1)^{k}(n\pi)^{2k}/(2k+1)!$, which apparently indicates that 
 its constant term is one. On the other hand, the summation results in 
$\sin n\pi/(n\pi)$, which vanishes for all $n=1,2,...$. 
  Such ambiguity causes  probability densities where the replica 
  method does not work \cite{d,t}.   

In this paper, we show that the constant term in $n$ 
of a physical quantity calculated in the replicated system 
equals the disordered average 
of the corresponding physical quantity in the original system
within finite-order perturbation. 
It is also found that the $n$-dependent terms 
 in the replicated system   
originate from the disorder correlation between the 
corresponding physical quantity and the free energy 
in the original system, which are nothing to do with 
the disorder average we want.  
The limit $n \rightarrow 0$
can be interpreted as realization of extracting the constant term. 

Our argument is also applicable to perturbative renormalization group, 
which includes functional renormalization group in perturbation \cite{f,wl}.  
We will see that a beta function calculated 
in perturbation becomes a polynomial in $n$ in the replicated system. 
The constant term in it is precisely equal to the beta function 
for the correlators characterizing probability density of quenched disorder.

It is pointed out in the literature \cite{bd}
that a replicated field theory describing the random-field Ising model (RFIM) yields 
one-loop correction to coupling constants singular in the limit $n \to 0$. 
A novel replica limit is  
proposed in order to remove the singularity.  
Since it provides beta function allowing breakdown of dimensional reduction 
for the RFIM near the upper critical dimensions 
contrary to
 the conventional limit \cite{aim}, it is important to know which the right way is. 
Although  the two replica limits can be compared with 
a direct computation not employing replicants \cite{ms}, 
the ambiguity could not  be resolved within the replica method so far 
because of missing mathematical basis of the limit. 
This issue will be treated as an example. 

\section{Expectation value}
Let us consider a field theory consisting of a single-component 
field $\phi$ and a quenched disorder $v$.
We write the thermal expectation value of a physical quantity $A[\phi]$, 
a functional of $\phi$, 
as 
\beq
  \bra A[\phi] \ket_{H[\phi; v]} \equiv 
  \frac{\int {\cal D}\phi \, A[\phi] \, e^{-H[\phi; v]}}{\int {\cal D}\phi \, e^{-H[\phi; v]}}, 
  \label{<>}
\eeq
where $H[\phi; v]$ is the Hamiltonian (times $1/kT$) of this model. 
We can put $H[0; v] = 0$ for an arbitrary $v$
 without loss of generality since the quotient (\ref{<>}) is independent of $H[0; v]$.  
 
The probability density for $v$  
is usually characterized by a given set of correlators among itself, 
which is denoted by $u$.  The average and the cumulants for the
quenched disorder are respectively described 
as $\rnd{}$ and $\kappa$. E.g., 
$
  \cum{}{v} = \rnd{v}, 
$
$
  \cum{}{v_1, v_2} = \rnd{\(v_1\, v_2\)} - \rnd{v_1} \rnd{v_2}. 
$

Our interest is to compute 
expectation values  such as $\rnd{\bra A[\phi] \ket_{H[\phi; v]}}$ by means of the replica method. 
Introducing $n$ identical systems, we define the replica partition function 
\beq
 \rZ \equiv \rnd Z^n 
 =  \rnd{ \int \prod_{\al=1}^n {\cal D} \phi_{\al} e^{ - \sum_{\al=1}^n H[\phi_\al ; v]}}. 
 \label{rz}
\eeq
The replica Hamiltonian is defined as 
\beqa
  - \rH[\bphi; u] &=& \log \rnd{e^{ - \sum_{\al=1}^n H[\phi_\al ; v]}}, 
\eeqa
where we have used the notation $ \bphi\equiv (\phi_1, ..., \phi_n)$. 
We restrict ourselves to the case where
the cumulant expansion for 
the right-hand side terminates at some finite number:
\beq
-\rH[\bphi;u] = \sum_{l=1}^{M}\frac{(-1)^l}{l!}
\cum{}{
\sum_{\alpha_1=1}^{n}H[\phi_{\alpha_1};v], \ldots, 
\sum_{\alpha_l=1}^{n}H[\phi_{\alpha_l};v]
}, 
\label{finite-num}
\eeq
which excludes introducing infinite number of replica indices. 
The physical quantity in the replicated system corresponding to 
$\rnd{\bra A[\phi] \ket_{H[\phi; v]}}$  
is  $\bra A[\phi_1] \ket_{\rH[\bphi; u]}$, where the subscript 1 is the replica index. 
It can be written as 
\beq
  \bra A[\phi_1] \ket_{\rH[\bphi; u]} = 
  \frac{ \rnd{\( \bra A[\phi] \ket_{H[\phi; v]} Z^n\)}}{\rnd{Z^n}}. 
  \label{<>rep2}
\eeq
  Now we evaluate the right-hand side in
 finite-order perturbation.
Letting $W=\log Z$,  the cumulant expansion  has the form of 
\beq
   \sum_{k=0}^\infty \frac{1}{k!} \, \cum{}{\bra A[\phi] \ket_{H[\phi; v]}, 
   \underbrace{n W, ... ,nW}_k }. 
   \label{cum-rep}
\eeq
We write the unperturbed Hamiltonian as $H_0[\phi; u]$.  
Here, the expression $H_0[\phi; u]$ instead of 
$H_0[\phi; v]$ implies that the disorder average for $H_0$ is already taken. 
Perturbation is defined as $V \equiv H[\phi;v] - H_0[\phi; u]$. 
The entries in the cumulant (\ref{cum-rep}) have the following perturbative series: 
\beqa
  && \bra A[\phi] \ket_{H[\phi; v]} = 
  \sum_{m=0}^\infty \frac{(-1)^m}{m!} \bra A[\phi]  ; \underbrace{V;...; V}_m \ket_{H_0 [\phi; u]} 
  \nn\\
  && W = \log Z_0 + \sum_{m=1}^\infty \frac{(-1)^m}{m!} 
  \bra   \underbrace{V;...; V}_m  \ket_{H_0 [\phi; u]}, 
  \label{pert-aw}
\eeqa
which are obviously independent of $n$. 
Here the semicolons in the angle brackets mean to 
take the connected part, e.g., $\bra A; B \ket = \bra A \, B\ket - \bra A \ket \bra B\ket$, 
and the partition function for $H_0$ is denoted by $Z_0$. 
The summations in (\ref{cum-rep}) and (\ref{pert-aw}) 
are truncated at a finite number under a finite-order perturbation, 
so that 
$\bra A[\phi_1] \ket_{\rH[\bphi;u]}$ is expressed in a polynomial in $n$. 
Let $[Y]_m$ be the perturbative expansion of an arbitrary quantity $Y$ 
up to and including the $m$-th order in $V$.  
Since the constant term of (\ref{cum-rep}) is given by the term with $k=0$, 
we have 
\beqa
  \proj{\left[ \bra A[\phi_1] \ket_{\rH[\bphi;u]} \right]_m} &=& 
  \cum{}{\left[\bra A[\phi] \ket_{H[\phi; v]} \right]_m} 
  \nn\\
  &=& 
  \rnd{\left[\bra A[\phi] \ket_{H[\phi; v]} \right]_m}. 
  \label{rel-ev}
\eeqa
It shows that the perturbative expansion for 
$\rnd{\left[\bra A[\phi] \ket_{H[\phi; v]} \right]_m}$ 
is given by the constant term in $n$ of the perturbative 
expansion for
$\bra A[\phi_1] \ket_{\rH[\bphi;u]}$. 
We also find from (\ref{cum-rep}) that the higher-order terms in $n$ 
give disorder correlation between $\bra A[\phi] \ket_{H[\phi; v]}$ and the 
free energy $W$, which should be generally removed. 
 We comment that substituting $W$ for $\bra A[\phi] \ket_{H[\phi; v]}$ yields
 the moments of the free energy from the higher-order terms
 \cite{gb,b-d}.

\section{Renormalization Group}
The perturbation becomes more powerful 
combined with the renormalization group (RG), 
which consists of a coarse graining and a rescaling \cite{wk}. 
The coarse graining means integrating over high-momentum components. 
Let the momentum space with a cutoff $\La$  be $\kt \equiv \{k :  0 \leq |k| \leq \La \}$. 
Introducing $L > 1 $, we divide $\kt$  into low- and high-momentum spaces defined as 
$\kl \equiv \{k :  0 \leq  |k| \leq L^{-1} \La \}$
and  $\kh \equiv \{k :  L^{-1} \La < |k| \leq \La \}$ respectively. 
The Hamiltonian $H[\phi; v]$ is also decomposed into 
$H^<[\phi; v]$ and $H^>[\phi; v]$, where $H^<[\phi; v]$ contains only 
the low-momentum component $\phi(p)$ ($p \in \kl$), and $H^>[\phi; v]$ denotes 
the remainders. 
Integrating over $\phi(q) \, (q \in \kh)$ in $Z$,  
we have the correction term $\de H[\phi, v]$ of the Hamiltonian generated as 
\beq
   \int \prod_{q \in \kh} d \phi(q) e^{-H^>[\phi, v]} =  e^{- \de H[\phi; v]}. 
\eeq 
The rescaling procedure is carried out introducing the renormalized field 
$\phi'(k) \equiv L^{-\theta} \phi(L^{-1}k)$ and the disorder $v'$. Here, 
a number $\theta$ and a renormalized disorder $v'$ is determined in 
such a way that a main part of the Hamiltonian remains the same form:
  \beq
    -H^< [\phi; v] -\de H[\phi; v] = - H[\phi'; v'] - \de v_0 + \cdots, 
    \label{ren_H}
  \eeq
where  $\de v_0$ means $\phi$-independent but $v$-dependent term, 
which should be added to the Hamiltonian because we put $H[0; v]=0$.   
The dots in the right-hand side represents irrelevant terms other than $\de v_0$, 
which can be neglected in low-energy physics. 
The same procedure leads to the RG transformation (RGT) for 
$\bra A[\phi] \ket_{H[\phi; v]}$ \cite{hl,l,m}, where $A[\phi]$ contains  
low-momentum components of $\phi$. 
Since the expectation value is independent of 
 $\de v_0$, the RGT  becomes
  \beq
   \bra A[\phi] \ket_{H[\phi; v]} = 
   \bra A[L^\theta \phi'] \ket_{H[\phi'; v']}. 
  \eeq
 Taking the average over quenched disorder, we get
   \beq
   \rnd{\bra A[\phi] \ket_{H[\phi; v]} }= 
      \rnd{ \bra A[L^\theta \phi'] \ket_{H[\phi'; v']}}. 
   \label{rgt-org}
  \eeq
Next, we calculate the counterpart in the replicated system. 
We apply to the RGT to $\rZ$ in (\ref{rz}) before taking the 
random average:
\beq
  \rZ = \rnd{\int \prod_{\al=1}^n {\cal D} \phi'_\al \, 
  e^{-\sum_{\al=1}^n H[\phi'_\al; v'] - n \de v_0} }, 
\eeq
which indicates that the renormalized replica Hamiltonian satisfies
\beq
  - \rH[\bphi' ; u'] = \log \, \rnd{%
    \, e^{- \sum_{\al = 1}^n H[\phi'_\al ; v'] - n \de v_0}
    }%
\label{Hbarla}
\eeq
up to the irrelevant terms. 
Thus, the RGT for the replicated system is written as 
\beq
  \bra A[\phi_1] \ket_{\rH[\bphi;u]} = 
  \bra A[L^{\theta} \phi'_1] \ket_{\rH[\bphi'; u']}. 
  \label{rgt-rep}
\eeq

Since there are no free replica 
sums in the right-hand side of (\ref{Hbarla}), 
$u'$ explicitly depends on $n$ through $n \, \de v_0$.
Furthermore, employing the cumulant expansion (\ref{finite-num}), 
the renormalized coupling constants are expressed as polynomials in $n$.
Combining (\ref{rel-ev}), (\ref{rgt-rep}) and (\ref{rgt-org}), 
we get 
\beq
 \proj{\left[\bra A[L^{\theta} \phi'_1] \ket_{\rH[\bphi';u']}\right]_m}=
 \rnd{\left[\bra A[L^{\theta} \phi'] \ket_{H[\phi'; v']}\right]_m}. 
 \label{rgt-proj}
\eeq
It clearly shows that the right-hand side of (\ref{rgt-org})  
can be computed from the RGT in the replicated system 
with the constant-term extraction.   
Note that $\proj$ removes $n$ dependence of $u'$. 
It implies that the beta function, which is defined by 
the  linear response of $u'$ under the infinitesimal change 
$L \rightarrow L + \de L$,  is independent of $n$.  

\section{Example}
Now we exhibit a couple of examples. 
One of the simplest example is the random-field Gaussian model 
in $d$ dimensions given by the following Hamiltonian 
\beq
  H_1[\phi; v] = \int_k \phi(k) v_1(k)  + 
  \frac{1}{2} \int_{k_1, k_2} v_2(k_1, k_2)\, \phi(k_1) \phi(k_2), 
\label{orig1}
\eeq
where $v_1$ and $v_2$ are quenched disorder. For simplicity, 
we ignore fluctuation of $v_2$ and fix 
\beq
  v_2\(k_1, k_2\) = \(k_1^2 + t \) \loc, 
  \label{cum00}
\eeq
where $\loc \equiv (2\pi)^d \de\(K\)$ with $K \equiv \sum_{i} k_i$, 
 while $v_1$ obeys 
\beq
  \cum{}{v_1(k_1), v_1(k_2)} = \De \, \loc. 
  \label{cum0}
\eeq
The subscript $k$ for the integral in (\ref{orig1}) means the measure $d^dk/(2\pi)^d$ on $\kt$.  
We regard the second term in (\ref{orig1})  as the unperturbed Hamiltonian $H_0$
 and the first term as perturbation $V$.  Since cumulants of $v_1$ higher than 
 (\ref{cum0}) vanish, the higher order terms with $m \geq 3$ in the first line of 
 (\ref{pert-aw}) vanish when we take the disorder average. Hence we can readily 
 derive the exact result
\beq
  \rnd{\bra \phi(k_1) \phi(k_2) \ket_{H_1[\phi; v]}} = \( G_0(k_1) + \De G_0(k_1)^2 \) \loc, 
  \label{E2pt}
\eeq
where $G_0(k) = 1/(k^2+t)$.  The same quantity is computed by the replica 
method as shown below. Employing (\ref{finite-num}) and (\ref{cum0}), 
the replica Hamiltonian to  (\ref{orig1}) is 
\beq
  \rH_1[\bphi ; u] = \sum_{\al, \be=1}^n \frac{1}{2} 
  \int_k \phi_\al (k) \( \(k^2 + t\) \de_{\al \be} - \De\) \phi_\be(-k). 
  \label{rH0}
\eeq
Perturbative expansion with respect to $\De$ shows that 
 \beqa
  && \left[ \bra \phi_1(k_1) \phi_1(k_2) \ket_{\rH_1[\bphi; u]} \right]_{2m} 
  \nn\\
  &=& \( G_0(k_1)  + \sum_{j=1}^{m} n^{j-1} \De^{j} G_0^{j+1}(k_1) \) \loc. 
  \label{rep2pt0}
 \eeqa
We see that the constant term in $n$ is identical with (\ref{E2pt}), as expected. 
On the other hand, 
if we take  
$m\rightarrow \infty$ in (\ref{rep2pt0}), we obtain the exact two-point function 
in  the replicated system:
\beqa
  && \bra \phi_1(k_1) \phi_1(k_2) \ket_{\rH_1[\bphi; u]} 
  \nn\\
 &=& G_0(k_1) \(1+ \frac{\De}{\(k_1^2 + t - n \De\)} \) \loc. 
  \label{rep2pt}
\eeqa
It is argued in Ref.\cite{bd} that, when $\phi^4$ coupling constants are taken into account, 
 the  limiting procedure putting $t = n \, \De$ and then $n \rightarrow 0$ in (\ref{rep2pt}) 
may generate a correction singular in $n$ to the  
coupling constants due to $n \, \De$ in the denominator. 
However, it is easily checked that the higher-order terms with $j \geq 1$ in (\ref{rep2pt0})
correspond to 
disorder correlation between $\bra \phi(k_1) \phi(k_2) \ket_{H[\phi; v]}$ and 
the free energy $W$, which should be removed. 
 
\begin{figure}
\begin{center}
\setlength{\unitlength}{1mm}
\begin{picture}(60, 23)(0,0)
     \put(0,0){ 
\includegraphics[width=60mm]{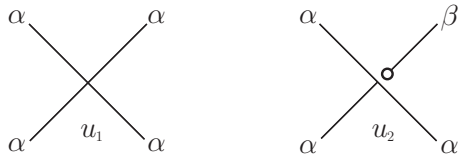}
		}
\end{picture}
\end{center}
\caption{Vertices appeared in the Hamiltonian (\ref{rH}), which are generated by 
$\cum{}{v_4}$ and $\cum{}{v_3, v_1}$ respectively. The open circle denotes $v_1$. 
 }
\label{fig_vertex}
\end{figure}

\begin{figure}
\begin{center}
\setlength{\unitlength}{1mm}
\begin{picture}(86, 15)(0,0)
     \put(0,0){ 
\includegraphics[width=86mm]{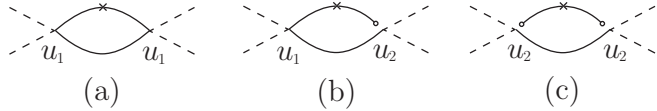}
		}
\end{picture}
\end{center}
\caption{The one-loop correction discussed in the main text. 
A low momentum component is shown in a broken line, 
while the solid line means $G_0(q), \, q \in \kh$. 
A cross on a solid line depicts $\Delta$ carrying 
$\Delta G_0(q)^2$, which is generated by $\cum{}{v_1, v_1}$.}
\label{fig_u1}
\end{figure}

Next, in order to include $\phi^4$ interactions discussed above,  
consider the following Hamiltonian inspired by \cite{hks, bd,ms}:
\beq
  H [\phi; v] = \sum_{l=1}^4 \frac{1}{l!} \,  \int_{k_1, ..., k_l} v_l(k_1, ..., k_l) \phi(k_1) \cdots \phi(k_l). 
  \label{orig2}
\eeq
Here, the quenched disorder has the following
 non-vanishing cumulants in addition to (\ref{cum00}) and (\ref{cum0}):
\beqa
   &&  \cum{}{v_4(k_1, k_2, k_3, k_4)} = u_1 \, \loc
  \nn\\
  &&\cum{}{v_3(k_1, k_2, k_3), v_1(k_4)} = -u_2 \, \loc. 
\eeqa 
They yield the following replica Hamiltonian employing (\ref{orig2}) and (\ref{finite-num}). 
\beq
  \rH[\bphi; u] = \rH_1[\bphi; u]  
  + \sum_{\al, \be=1}^n  \sigma_{\al \be}
  \( \frac{u_1}{4!}  \de_{\al \be} + 
  \frac{u_2}{3!} 
  \),
\label{rH}
\eeq
where $\rH_1$ is given in (\ref{rH0}) and 
\beq
  \sigma_{\al \be}  
  \equiv \int_{k_1, ..., k_4} \loc  \phi_\al (k_1) \phi_\al (k_2)
  \phi_\al (k_3) \phi_\be (k_4). 
\eeq
Here we focus on the one-loop corrections to $u_1$
having one $\Delta$. We treat $\rH_1[\bphi; u]$ as the unperturbed Hamiltonian, 
while the remaining terms in (\ref{rH}) the perturbation graphically represented as 
Fig. \ref{fig_vertex}.  
 As we discussed in the above 
example, the free propagator is (\ref{rep2pt}) with $n\Delta$ removed. 
The $n$-dependent, one-loop diagrams having one $\Delta$ 
for $u'_1$ are presented in Fig.\ref{fig_u1}, which are calculated as \cite{bd}
\beq
  \frac{\Delta}{4!} \( 3 u_1^2 + 6 n u_1 u_2 + 3 n^2 u_2^2 \)  \int_{q \in \kh} G_0(q)^3.    
  \label{deu_1}
\eeq
The novel limiting procedure proposed in \cite{bd} is $n\rightarrow 0$ 
with $g_2 \equiv n u_2$ fixed,  (\ref{deu_1}) yields  
the following beta function $\beta_1$ for  $g_1 \equiv \Delta u_1$:
\beq
  \beta_1 = (6-d) g_1 -  (3 g_1^2 + 6 \Delta g_1 g_2 + 3 \Delta^2 g_2^2). 
  \label{beta1}
\eeq
It means that $g_2$ can affect flow of $g_1$. In Ref. \cite{bd},  a similar mechanism is argued 
in the context of failure of dimensional reduction. 

\begin{figure}
\begin{center}
\setlength{\unitlength}{1mm}
\begin{picture}(86, 25)(0,0)
     \put(0,0){ 
\includegraphics[width=86mm]{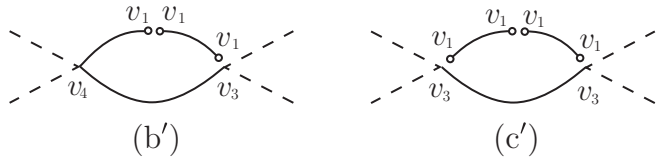}
		}
\end{picture}
\end{center}
\caption{Diagrams corresponding to (b) and (c) in Fig. \ref{fig_u1} before taking random average}
\label{fig_u1b}
\end{figure}

Contrary to this procedure, our general 
argument indicates that the second and the third terms in 
(\ref{deu_1}) should be removed 
because they are generated by the disordered  average in (\ref{Hbarla}) containing 
$n  \de v_0$.  In fact, the diagrams in Fig.\ref{fig_u1b}, 
which corresponds to those in Fig.\ref{fig_u1} before taking the random average,  have
$n v_1(q) G_0(q) v_1(-q)$ contained in perturbative expansion in 
$n \de v_0$.   Thus the beta function for $g_1$ according to our argument becomes 
\beq
  \bar{\beta}_1 = (6-d) g_1 -  3 g_1^2. 
  \label{beta11}
\eeq
Here, $g_2$ does not affect flow of $g_1$ within the one-loop correction, which 
leads to dimensional reduction. 
%
Although we have demonstrated the simpler model, 
a tedious but straightforward computation in the original model \cite{bd}
 shows that all the
terms depending on $n$ and remaining in the novel limiting procedure 
are caused by disorder correlations with $n \de v_0$. These terms drastically change 
stability of fixed points and lead to breakdown of dimensional reduction \cite{bd}. 
However, in the case when the higher order terms in $n$ are dropped in the beta function, 
this scenario does not happen and dimensional reduction does occur \cite{ms,tt}. 
Although it is still debated 
whether dimensional reduction occurs or not near the upper critical 
dimensions \cite{f,bd,aim,ms,hks,tt,s,gaahs,fe,tt2}, 
the seeming  
ambiguity of the limiting procedure $n \rightarrow 0$ in perturbation 
cannot explain its breakdown. 

\section{Summary and Discussion}
We have shown that a physical quantity in a quenched disordered system 
calculated in a finite-order perturbation
can be exactly derived by the replica method 
where the limit $n \rightarrow 0$ 
means to extract the constant term in $n$. 
The $n$-dependent terms give disorder correlations 
between the physical quantity we want to compute and the free energy, 
which should be removed.  
In this sense, the limiting procedure is uniquely determined as long as 
the physical quantity is represented in a polynomial in $n$.  Our claim can provide 
 a part of mathematical basis for various results by perturbative RG 
with the replica method where the limit is properly understood. 

\acknowledgments
The authors would like to thank E. Kanzieper for 
informing us of his work \cite{k}.


\begin{thebibliography}{10}
\bibitem{ea} S. F. Edwards and P. W. Anderson, 
J. Phys. F: Metal Phys., {\bf 5}, 965 (1975). 
\bibitem{cfs} F. Constantinescu, J. Fr\"olich and T. Spencer, 
J. Stat. Phys, {\bf 34}, 571 (1984)
\bibitem{kmp} A. Klein, F. Martinelli and J. Perez, Commun. Math. Phys, 
{\bf 106},  623 (1986)
\bibitem{k} E. Kanzieper, Phys. Rev. Lett. {\bf 89} (2002) 250201; 
in  {\it Frontiers in Field Theory}, edited by O. Kovras, Ch. 3,  pp. 23 -- 51
 (Nova Science Publishers, NY 2005); 
 V. A. Osipov and E. Kanzieper, Phys. Rev. Lett. {\bf 99} (2007) 050602
\bibitem{b-d} \'E. Brunet and B. Derrida, Phys. Rev. E {\bf 61} 6789 (2000). 
\bibitem{d} B. Derrida,  Phys. Rev. B {\bf 24} 2613 (1981). 
\bibitem{t} T. Tanaka, Interdisciplinary Information Sciences, {\bf 13} 17 (2007)
available online. 
\bibitem{f}
D. S. Fisher, Phys. Rev. B {\bf 31},   7233 (1985).
\bibitem{wl}
K. J. Wiese, P. LeDoussal, Markov Processes Relat. Fields {\bf 13}  777 (2007), 
for a review. 
\bibitem{bd}
E. Br\'ezin and C. De Dominicis, Europhys. Lett. {\bf 44},  13  (1998).
\bibitem{aim}
A.  Aharony, Y. Imry and S. K. Ma,  Phys. Rev. Lett. {\bf 37},  1364 (1976) \\
A. P. Young, J. Phys. {\bf C10},  L275  (1977)\\
G. Parisi and N. Sourlas, Phys. Rev. Lett. {\bf 43}, 744 (1979).
\bibitem{ms}
H. Mukaida and Y. Sakamoto, Int. J. Mod. Phys. B. {\bf 18}, 919 (2004). 
\bibitem{gb}
D. A. Gorokhov and G. Blatter, Phys. Rev. Lett.  {\bf 82},  2705 (1999).
\bibitem{wk}
K. G. Wilson and J. Kogut, Phys. Rep. {\bf 12 C}, 75,  for a review.
\bibitem{hl}
 A. B. Harris and T. C. Lubensky, Phys. Rev. Lett.  {\bf 23},  1540 (1974).
\bibitem{l}
T. C. Lubensky, Phys. Rev. B {\bf 11},  3573 (1975).
 \bibitem{m}
S. K. Ma {\it Modern Theory of Critical Phenomena}, Chapter X, Benjamin (1976) 
\bibitem{hks}
A. Houghton, A. Khurana and F. J. Seco, 
Phys. Rev. B  {\bf 34},   1700 (1986)
\bibitem{tt} G. Tarjus and M. Tissier, Phys. Rev. B {\bf 78}, 024203 (2008).
\bibitem{s} 
M. Schwartz, J. Phys. C  {\bf 18},  135 (1985).
\bibitem{gaahs}
M. Gofman, J. Adler, A. Aharony, A.B. Harris, and M. Schwartz, Phys. Rev. B {\bf 53},   6362 (1996)
\bibitem{fe} D. E. Feldman, Phys. Rev. Lett. 88, 177202 (2002).
\bibitem{tt2} M. Tissier and G. Tarjus, Phys. Rev. B {\bf 78}, 024204 (2008).
\end{thebibliography}
\end{document}